\documentclass[aps,prb,twocolumn,superscriptaddress,longbibliography,citeautoscript]{revtex4-1}
\pdfoutput=1

\usepackage{amsmath}
\usepackage{amssymb}	
\usepackage{graphicx} 
\usepackage{color}
\usepackage[normalem]{ulem}
\usepackage[protrusion=true,expansion=true]{microtype} 
\usepackage[english]{babel}

\usepackage{hyperref} 
\hypersetup{
	pdftoolbar=true,        		 
	pdfmenubar=true,     		 
	pdffitwindow=false,     	 
	pdfstartview={FitH}, 
	pdftitle={Kosterlitz-Thouless scaling at many-body localization phase transitions},  
	pdfauthor={Dumitrescu, Goremykina, Parameswaran, Serbyn, Vasseur},     		 
	pdfsubject={},   			 
	pdfcreator={},   	        	   	
	pdfproducer={}, 		
	pdfnewwindow=true,      	
	colorlinks=true,       		
	linkcolor=black,          	
	citecolor=blue,        		
	filecolor=magenta,    		
	urlcolor=blue          		
}

\usepackage{xcolor}

\begin{document}

\title{Kosterlitz-Thouless scaling at many-body localization phase transitions}
\author{Philipp T.~Dumitrescu}
\email{pdumitrescu@flatironinstitute.org}
\affiliation{Center for Computational Quantum Physics, Flatiron Institute, 162 5th Avenue, New York, NY 10010, USA}
\author{Anna Goremykina}
\affiliation{D\'{e}partement de Physique Th\'{e}orique, Universit\'{e} de Gen\`{e}ve, CH-1211 Gen\`{e}ve 4, Switzerland}
\affiliation{IST Austria, Am Campus 1, 3400 Klosterneuburg, Austria}
\author{Siddharth A.~Parameswaran}
\affiliation{Rudolf Peierls Centre for Theoretical Physics, Clarendon Laboratory,  University of Oxford, Oxford OX1 3PU, UK}
\author{Maksym Serbyn}
\affiliation{IST Austria, Am Campus 1, 3400 Klosterneuburg, Austria}
\author{Romain Vasseur}
\affiliation{Department of Physics, University of Massachusetts, Amherst, Massachusetts 01003, USA}

\date{\today}

\begin{abstract}
We propose a scaling theory for the many-body localization (MBL) phase transition in one dimension, building on the idea that it proceeds via a `quantum avalanche'. We argue that the critical properties can be captured at a coarse-grained level by a Kosterlitz-Thouless (KT) renormalization group (RG) flow. On phenomenological grounds, we identify the scaling variables as the density of thermal regions and the lengthscale that controls the decay of typical matrix elements. Within this KT picture, the MBL phase is a line of fixed points that terminates at the delocalization transition. We discuss two possible scenarios distinguished by the distribution of rare, fractal thermal inclusions within the MBL phase. In the first scenario, these regions have a stretched exponential distribution in the MBL phase. In the second scenario, the near-critical MBL phase hosts rare thermal regions that are power-law distributed in size. This points to the existence of a second transition within the MBL phase, at which these power-laws change to the stretched exponential form expected at strong disorder. 
We numerically simulate two different phenomenological RGs previously proposed to describe the MBL transition. Both RGs display a universal power-law length distribution of thermal regions at the transition with a critical exponent $\alpha_c=2$, and continuously varying exponents in the MBL phase consistent with the KT picture.
\end{abstract}
\maketitle

\section{Introduction}   

The question of how many-particle systems reach thermal equilibrium under their intrinsic dynamics has attracted renewed attention over the past decade, as experiments with a variety of synthetic quantum systems --- ultracold atoms \cite{RevModPhys.80.885, schreiber2015observation}, trapped ions \cite{leibfried2003single-ions, duan2010network-ions, Blatt2012SimulationsIons}, nitrogen-vacancy centers \cite{Doherty2013NVreview, Schirhagl2014NVreview} and superconducting qubits \cite{kelly2015state, roushan2017spectroscopic} --- can access new non-equilibrium regimes. Such systems are naturally isolated from an external environment and therefore evolve under reversible unitary dynamics. Their thermalization is best viewed as the `scrambling' of quantum information as it is transferred to non-local degrees of freedom, becoming essentially inaccessible to local measurements. The eigenstate thermalization hypothesis (ETH) provides a natural explanation of such dynamics via properties of individual eigenstates~\cite{PhysRevA.43.2046, PhysRevE.50.888}. ETH suggests that small parts of the quantum system experience the remaining degrees of freedom as a thermal bath. This relies essentially on the effective exchange of quantum information and transport of conserved quantities between different parts of the system~\cite{Polkovnikov-rev}.

In  many-body localized (MBL) phases, strong disorder arrests {the} efficient exchange of information, leading to a breakdown of thermalization even in interacting systems~\cite{PhysRev.109.1492, FleishmanAnderson, Altshuler1997QuasiparticleLifetime, Gornyi, BAA, PhysRevB.75.155111, PalHuse, ARCMP,VasseurMoore2016MBLReview,AbaninRev}. MBL phases have attracted much interest, as their lack of thermalization allows quantum correlations to persist at energy densities where they would be washed out in equilibrium. Highly excited eigenstates of MBL systems satisfy an area-law scaling for the entanglement entropy of subsystems, resembling ground states of local Hamiltonians~\cite{Bella}. In contrast, their ETH counterparts show a much higher volume-law scaling of entanglement~\cite{Polkovnikov-rev}. In one dimensional (1D) disordered systems, our focus here, the existence of MBL has been rigorously established~\cite{Imbrie16,ImbriePRL}. The proof of the existence of MBL sheds light on its underlying structure, as it proceeds by constructing an extensive set of (quasi-)local integrals of motion (LIOMs) or `l-bits', that commute with the Hamiltonian and are therefore conserved under time evolution. Much of the physics of the MBL phase --- lack of ergodicity, logarithmic dephasing time, and stability to weak perturbations --- can be understood in terms of these conserved quantities~\cite{Serbyn13-2,Huse13,Serbyn13-1}. This description in terms of emergent conserved quantities imparts an integrable structure to the MBL phase distinct from, and more stable than, that of translationally-invariant quantum integrable systems described by the Bethe ansatz~\cite{AbaninRev}.

Decreasing disorder or increasing interaction strength can eventually lead to a breakdown of many-body localization and restoration of ergodicity. The many-body (de)localization phase transition separates  MBL and thermalizing phases. This transition describes a breakdown of equilibrium statistical mechanics and has many features that distinguish it from phase transitions usually encountered in either quantum or equilibrium classical systems. Despite focused efforts in studying this transition~\cite{PalHuse,VHA,PVP,DVP,THMdR_meanfield,TMdR_numerics,ZhangHuse,GVS,abanincriterion,KjallIsing,Luitz,PhysRevX.7.021013,PhysRevLett.119.075702, PhysRevLett.121.206601}, the combination of finite energy density, interactions and disorder remains a challenge to all known theoretical and computational techniques. Much of the insight into the existence of the MBL phase and transition have been obtained by exact diagonalization studies, performed on small system sizes ($L \lesssim 30$). However, scaling exponents extracted from such studies~\cite{KjallIsing,Luitz} violate general bounds required by the self-consistency of the transition~\cite{Harris,Chayes,Chandran2015}. This suggests that numerical studies are subject to strong finite-size effects, limiting their reliability to deep inside the phases, where correlation lengths are short.

In light of these challenges, theoretical efforts have focused on phenomenological approaches that abandon a microscopically faithful treatment in favor of a coarse-grained description~\cite{VHA,PVP,DVP, THMdR_meanfield,TMdR_numerics, ZhangHuse, GVS}. These approaches were designed to identify the physical mechanism that drives the transition and build an effective model which could then be solved numerically for large system sizes.  Nonetheless, both the choice of a consistent model and the interpretation of its results in the context of the MBL transition have presented challenges. Despite being based on the same philosophy of coarse graining many-body resonances in a strong disorder approach, various proposed renormalization group (RG) approaches differ significantly in their procedures and their link to the microscopic physics. Thus,  a consistent picture of the critical point is missing. 

In this paper, we formulate a unifying scaling theory for the MBL transition that has a Kosterlitz-Thouless form. We show that the basic features of KT scaling emerge from a phenomenological description of the proliferation of  `quantum avalanches'~\cite{deRoeckHuveneersavalanche}  that drive the MBL transition. As such, this picture is independent of any specific microscopic model.  Specifically, we show that the avalanche process combined with a natural choice of scaling variables immediately leads to KT critical behavior.
The KT picture implies that the MBL critical point is the terminus of a line of RG fixed points characterized by an exactly marginal scaling variable. We discuss how this picture resolves many shortcomings of previous descriptions. However, it also raises questions about the physics beyond avalanches in the MBL phase away from the transition. Thus, in Section~\ref{sec:transitions}, we propose two distinct scenarios for the MBL phase distinguished by how the KT scaling is linked to a Griffiths description of the fractal rare thermal regions. 

Several  numerically tractable effective models have been previously proposed as a route to accessing scaling properties of the MBL transition. These include models designed to capture quantum avalanche processes~\cite{DVP,TMdR_numerics}, as well as ones where avalanches were not an apparent feature~\cite{VHA}. 
However, the transitions studied in those works were not identified as KT-like; this is perhaps unsurprising in light of the notorious difficulties in observing KT scaling even in classical equilibrium models. In light of the KT picture, we now revisit two of these models, in both cases dramatically increasing the available statistics or system sizes compared to previous studies.
 In Section~\ref{sec:numerics}, we reconsider the cluster RG  of Ref.~\onlinecite{DVP}, referred to as `DVP' in what follows. By analysing thermal distributions that are a direct output of this scheme we find an algebraic structure of thermal resonances in the MBL phase -- strong evidence for the KT flow.  In Section~\ref{sec:numerics2}, we implement the block RG of Ref.~\onlinecite{VHA}, referred to as `VHA' in the following, and also find results consistent with the KT picture. We comment on how the results of Sections~\ref{sec:numerics} and~\ref{sec:numerics2} may be accommodated within the two scenarios proposed in Section~\ref{sec:transitions}. Finally, we close in Section~\ref{sec:disc} with a summary of our main results and an overview of new directions in the  study of MBL transitions opened by the present work.

\section{\label{sec:avalanche}Phenomenological Argument for Kosterlitz-Thouless Scaling} 

\subsection{Many-body delocalization via quantum avalanches \label{sec:avalanchereview}}

\begin{figure}[b]
\includegraphics[width=\columnwidth]{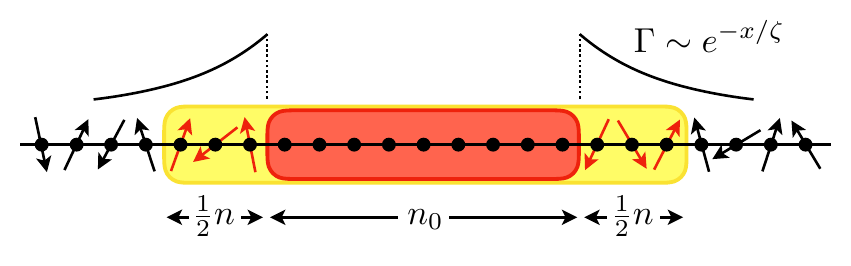}
\caption{\label{fig:avalanche}Quantum Avalanche~\cite{deRoeckHuveneersavalanche}. A thermal inclusion  initially consisting of $n_0$ spins (red region) is in contact with a set of l-bits (arrows). The inclusion thermalizes $n$ l-bits (red arrows) and thereby expands to a size $n_0+n$ (yellow region) while retaining its featureless ETH character. The effective matrix element to add the $(n+1)^{\text{th}}$ l-bit decays exponentially from the boundary of the original inclusion.}
\end{figure} 

Assuming a direct transition between the MBL and delocalized phases, at the transition, eigenstates undergo a complete rearrangement as the entanglement jumps abruptly from area-law to volume-law~\cite{VHA,PVP, DVP, Grover14,PhysRevX.7.021013}. This is quite unlike conventional critical points, which are driven by fluctuations of a locally defined order parameter. Numerical studies of the transition show strong asymmetry: a strongly resonant thermal block can thermalize a localized region far more effectively than a localized region can arrest the growth of the thermal block~\cite{luitz2017smallbath}. 

The asymmetry between thermalization and localization was formulated as an `avalanche' process that we briefly review  following Ref.~\onlinecite{deRoeckHuveneersavalanche}. Imagine a rare thermal region of $n_0$ spins (a `bubble') in an otherwise localized spin-${1}/{2}$ chain. Such a rare thermal inclusion is unavoidable in a generic system, with uncorrelated disorder. It will act as a small bath and will increase its size by thermalizing spins peripheral to it. Let us assume that the bubble has absorbed a number $n\gg1$ of l-bits to grow to a new size $n_0 + n$, but is still described by random matrix theory and thus {remains} featureless. Further growth of the bubble depends on the matrix element for flipping an l-bit at distance $n/2$ from the new edge (see Fig.~\ref{fig:avalanche}). This is asymptotically given by $\Gamma \sim  e^{-n/(2 \zeta)}/\sqrt{2^{n_0 + n}}$, where $2^{n_0 + n}$ is the dimension of the bubble Hilbert space and $\zeta$ characterizes the exponential decay of typical matrix elements with distance. This matrix element should be compared to the level spacing of the bubble $\delta \sim 2^{-(n_0 + n)}$:
\begin{equation}\label{eq:avalancheg}
g = \frac{\Gamma}{\delta}  \sim \exp\left(-\frac{n}{2\zeta} + \frac{\ln 2}{2}(n+n_0) \right).
\end{equation}
If $\zeta > \zeta_c =(\ln 2)^{-1}$, the matrix element falls off slower than the level spacing and $g \gg 1$ for $n\gg 1$. 
This leads to an ``avalanche'' process where the initial thermal inclusion will be able to repeatably absorb l-bits and grow until it thermalizes the whole system.  We note that this simple picture relies on the assumption that the growing inclusion obeys ETH at all steps~\cite{deRoeckHuveneersavalanche}. This avalanche process is supported by exact diagonalization studies on toy models that incorporate `random-matrix-type' inclusions~\cite{luitz2017smallbath, potirniche2018stability}; however it remains to be tested for fully microscopic lattice models.

We emphasize that the growth of ETH bubbles by absorbing spins is controlled by the effective interaction matrix elements of these resonances, which have to be carefully considered. 
Tracking the evolution of the effective coupling strengths and the degree of instability to thermalization at long distances is the purview of RG methods, to which we now turn.

\subsection{Kosterlitz-Thouless scaling \label{sec:avalancheKT}}

We now argue that the basic ingredients of the avalanche discussed above give rise to a Kosterlitz-Thouless scaling at the MBL transition, with minimal additional assumptions. Already implicit in the avalanche discussion is a degree of coarse graining, due to the presence of fully thermal regions at intermediate scales that arise out of microscopic configurations. We shall proceed with this picture, which we emphasize is not tied to any specific model, and will comment further on its validity below.

Given the presence of thermal regions that grow to drive the delocalization transition, it is natural to work with variables that capture the distributions of individual locally thermal blocks and their effectiveness in thermalizing neighboring regions. First, we identify the average density of thermal blocks $\rho(\ell)$ as a scaling variable. Here $\Lambda =  \Lambda_0 e^{-\ell}$ is the RG scale at which we are probing the system and $\Lambda_0 \sim 1 / a$ is the cutoff scale set by the lattice spacing $a$. As the second scaling variable, we identify the length scale $\zeta(\ell)$ that governs the effective matrix element ${\Gamma}(\ell) \sim e^{-x/\zeta(\ell)}$  at a distance $x$ from the boundary of a thermal block. These scaling variables control the distributions of physical observables, that are themselves broad at criticality due to the strong randomness inherent to the MBL transition.

It remains to deduce the RG equations that describe how $\rho,\zeta$, transform as the RG flows to longer length scales. Following the avalanche scenario outlined above, we first demand that at any scale, the density of thermal regions $\rho$ increases (decreases) under the RG if the typical localization length $\zeta$ at that scale is larger (smaller) than some critical value $\zeta_c$, corresponding to the onset (absence) of avalanche processes. The simplest flow equation consistent with this  is
\begin{equation}\label{eq:KT1}
\frac{ d\rho}{d\ell} = b \rho\left(\zeta-\zeta_c\right) +\ldots,
\end{equation}
where $b\sim O(1)$ is a positive constant, and the ellipsis denote higher order terms in $\rho$ and $(\zeta - \zeta_c)$. In RG language, Eq.~\eqref{eq:KT1} indicates that thermal resonances are relevant if $\zeta > \zeta_c$; they proliferate even when they are asymptotically rare.  Accordingly, we set $\zeta^{-1}_c = \ln 2$~\cite{THMdR_meanfield}.

Next, we consider the effect of the resonant regions on the matrix elements. Intuitively, $\zeta$ should be renormalized by thermal spots, and must grow under coarse-graining. Thermal inclusions can `shortcut' the exponential decay of matrix elements in the MBL phase, leading to an effective localization length $\zeta$ that is larger than the microscopic one. To leading order, the simplest RG equation consistent with this reads
\begin{equation}\label{eq:KT2}
\frac{ d\zeta^{-1}}{d\ell}  =  -c \rho\zeta^{-1} +\dots,
\end{equation}
where $c$ is a positive constant, and we assumed that $\zeta$ is not renormalized in the absence of thermal regions ($\rho=0$). A similar equation can be derived from the `law of halted decay' of Ref.~\onlinecite{THMdR_meanfield}.  

\begin{figure}
\includegraphics[width=\columnwidth]{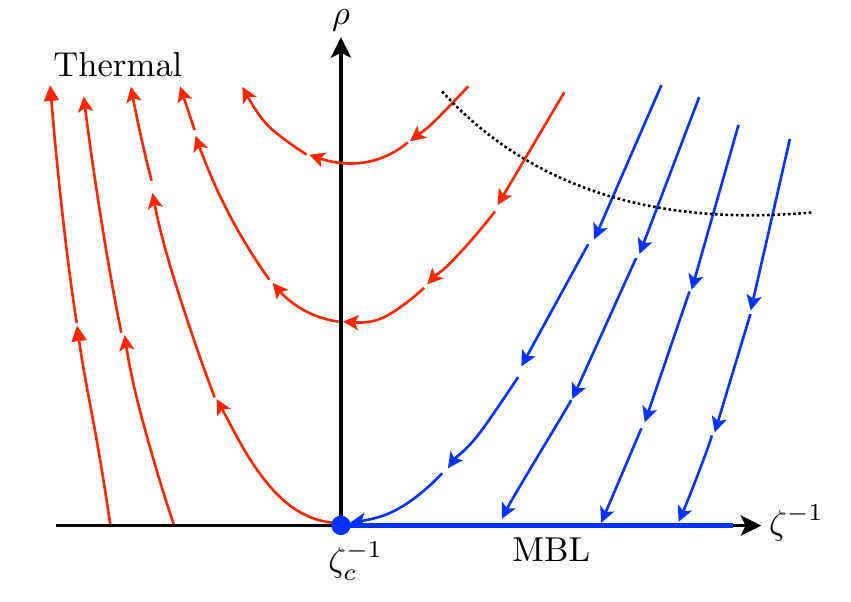}
\caption{\label{fig:KTflows} Kosterlitz-Thouless RG flow obtained by integrating Eqs. (\ref{eq:KT1}) and (\ref{eq:KT2}). The MBL phase corresponds to a line of fixed points with $\rho=0$ for $\zeta<\zeta_c$. For $\zeta>\zeta_c$ even an infinitesimally small bare density of resonances grows under RG, driving the flow to the thermal phase. The dotted line denotes a schematic line of microscopic parameters, tuned e.g. by decreasing disorder strength $W$. Note that many RG  trajectories initially approach the MBL fixed line even if they eventually flow to the thermal phase; this non-monotonicity naturally explains why numerical simulations often overestimate the extent of the MBL phase.}
\end{figure}

Equations (\ref{eq:KT1}) and (\ref{eq:KT2}) yield RG flows of the Kosterlitz-Thouless form (Fig.~\ref{fig:KTflows}), whose physical interpretation we now discuss. For $\zeta^{-1} >\zeta_c^{-1}$, {these RG equations} admit a line of  stable fixed points corresponding to the MBL phase, where the effective density of the thermal regions vanishes at long wavelengths, i.e. $\rho_\infty \equiv \rho(\ell \to \infty) \rightarrow 0$. Points along this line may be parameterized by the fixed-point value of the typical localization length $\zeta_{\infty}=\zeta(\ell \to \infty)$. For $\zeta^{-1}<\zeta_c^{-1}$, $\rho$ is relevant and flows to infinity, indicating the proliferation of thermal spots: this is the delocalized, thermal phase. At the critical point, $\zeta_\infty^{-1}$ jumps discontinuously, analogous to the stiffness discontinuity in the usual XY transition~\cite{KosterlitzThouless}. Assuming {that the disorder strength $W$ is} the parameter that tunes across the transition,  $\zeta^{-1}$ evolves as
\begin{equation}\label{eq:KTjump}
\zeta_\infty^{-1} =  {\zeta_c}^{-1} + c_1\sqrt{W-W_c} + \dots, 
\end{equation}
for $W>W_c$, whereas it is formally 0 in the delocalized phase. We emphasize that $\zeta_c^{-1} = \ln 2$ is a universal number in this scenario, which does not depend on microscopic details other than the dimension of the on-site Hilbert space. In general, it is given by the entropy density of the system at infinite effective temperature --- corresponding to the level spacing in the middle of the many-body spectrum.

Whereas the typical localization length $\zeta$ remains finite up until the transition, finite-size scaling is controlled by an emergent, diverging length scale
\begin{equation}\label{eq:KTcrossover}
\xi_{\pm} \sim e^{{c_\pm}/{\sqrt{|W-W_c|}}},
\end{equation}
 where $c_\pm>0$ {is non-universal and may be different for the two sides of the transition}. This corresponds to a correlation length exponent $\nu =\infty$, which is consistent with general bounds that require $\nu \geq 2$ in $d=1$~\cite{Harris,Chayes,Chandran2015}. 
 
The two-parameter phase diagram governed by variables $\rho, \zeta$ and the resulting KT picture of the MBL transition is the main result of the present work. It is consistent with a recent exactly solvable RG scheme for the transition~\cite{GVS}. Before discussing the physical implications of our results, let us discuss the potential relation between this KT scenario and other theories of the MBL transition. Numerical simulations of phenomenological RG approaches designed to capture the critical behavior suggest a finite value of the {correlation length} critical exponent around $\nu \approx 3.5$~\cite{VHA,PVP,DVP}. We note that the finite values of $\nu$ obtained from finite-size scaling collapses do not rule out the KT scenario, since KT transitions exhibit significant, logarithmic corrections to finite-size scaling. We reexamine the RG approach of Ref.~\onlinecite{DVP} in Section~\ref{sec:numerics} and of Ref.~\onlinecite{VHA} in Section~\ref{sec:numerics2}, finding support for the KT scenario. 

In a different direction, previous microscopic numerical studies~\cite{Luitz} have yielded values of critical exponent $\nu<1$  that are inconsistent with the general bound $\nu \geq 2$~\cite{Harris,Chayes,Chandran2015}. This may be potentially reconciled with the KT scenario by observing that flow in Fig.~\ref{fig:KTflows} leads to  \emph{overestimation} of the MBL phase. RG trajectories shown by red lines in Fig.~\ref{fig:KTflows} (corresponding to $\zeta_\infty > \zeta_c$) initially flow towards the MBL phase but  ultimately end  in the thermal phase. When cut off by small finite sizes, such flows would appear to correspond to the localized phase. This behavior has been generally noted~\cite{VHA, abanincriterion,DevakulSingh, Gopa18, Doggen2018mblTDVP}.
 
 Our KT scenario relies on two main assumptions: namely, that (i) the transition is driven by a quantum avalanche~\cite{deRoeckHuveneersavalanche}, so that a single resonant region can thermalize the whole system, and (ii) the critical behavior can be captured using two scaling variables $\rho(\ell)$ and $\zeta(\ell)$. The avalanche scenario is well-established by now, and underlies different recent phenomenogical RG schemes for the transition~\cite{DVP,THMdR_meanfield,TMdR_numerics}. The second assumption of only two scaling variables is more severe: it is motivated by the conceptual simplicity of the resulting KT picture. Specifically, the proliferation of thermal blocks in a quantum avalanche intuitively resembles vortex unbinding at the KT transition: both describe how a perturbatively stable phase responds to nonperturbative inclusions, respectively resonances or vortices. This emphasizes the importance of describing the critical behavior from the perspective of the stability of the localized phase: it is more aptly viewed as a many-body {\it de}localization transition. The KT picture also makes various previously puzzling aspects of the transition much easier to rationalize. For instance, KT scaling readily accommodates the first-order-like discontinuities characteristic of some quantities at the MBL transition~\cite{PhysRevX.7.021013, DVP}. Second, the KT picture applied to the MBL transition predicts that the critical point is a smooth continuation of the MBL phase; this is consistent with Refs.~\onlinecite{TMdR_numerics,THMdR_meanfield, GVS}. Third, as mentioned above, the logarithmic finite-size scaling expected for a KT transition  naturally explains  the slow convergence of numerical studies of the MBL transition to the thermodynamic limit. These individual pieces of evidence, although circumstantial, provide {further} support to the KT hypothesis. 

\section{\label{sec:transitions} Consequences for the MBL phase and Griffiths regions} 

In the KT scenario for the MBL transition proposed above,  the MBL phase is identified as a line of fixed points where $\zeta$ varies continuously. Thus, it should be possible to smoothly connect the KT scenario with the behavior of rare (Griffiths) thermal inclusions within the MBL phase. In this section, we show that KT scaling is compatible with the Griffiths picture. We propose two possible scenarios connecting the effective typical localization length $\zeta$ to the distribution of the length of thermal inclusions in the MBL phase. This also illustrates how the RG equations  (\ref{eq:KT1})-(\ref{eq:KT2}) are related to infinite-randomness physics.

\begin{figure}
\includegraphics[width=\columnwidth]{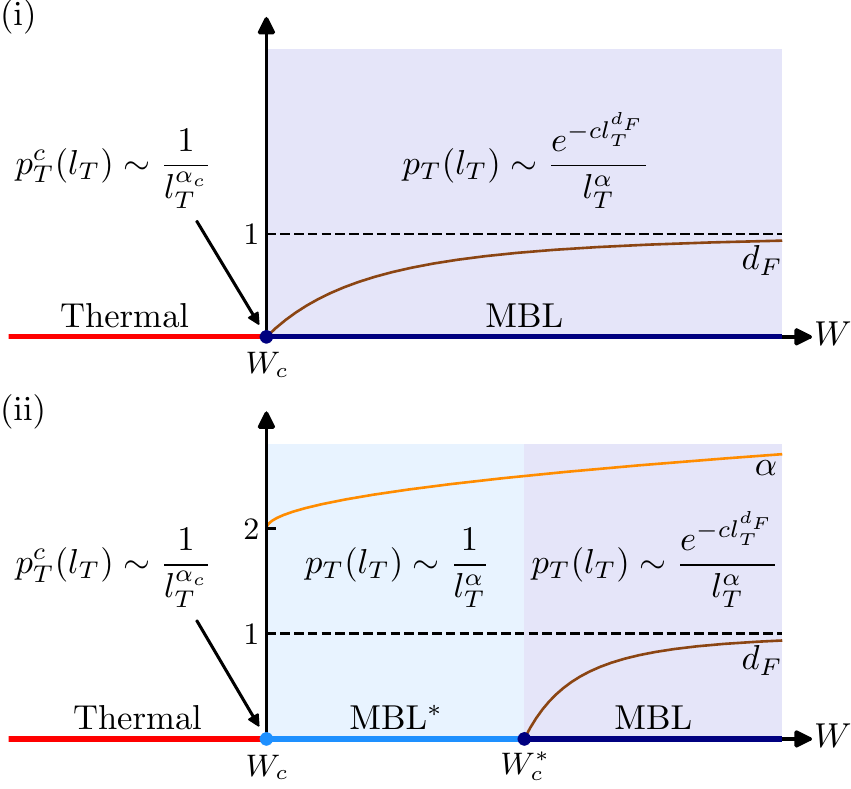}
\caption{\label{fig:twotrans} Two scenarios for the {evolution of the Griffiths regions} from critical to strong disorder. 
In scenario (i), thermal blocks have power-law distribution only at criticality, $W= W_c$ with $p^c_T(l_T) \sim l_T^{-\alpha_c}$. All other points in the MBL phase have stretched exponential thermal block distributions, characterized by an evolving non-zero fractal dimension $d_F>0$ {of the rare regions}, saturating deep in the MBL phase to $d_F=1$. In the `two transition' scenario (ii)  {thermal blocks are power-law distributed across a critical MBL* phase, until} a second critical disorder strength $W^*_c$ is reached. At the second transition the thermal block distribution changes from power-law ($\text{MBL}^*$) to stretched exponential (MBL), with the fractal dimension again continuously approaching $d_F=1$. }
\end{figure}

\subsection{Nature of Griffiths regions}

The physics of the MBL phase near the transition is dominated by rare (Griffiths) thermal inclusions that are locally on the `wrong side' of the transition~\cite{PhysRevLett.23.17, Demler14,VHA,PVP,Reichman15,mbmott,PhysRevB.93.134206,luitz2016,Znidaric16,Agarwal2017rareregion}. Such rare regions are unavoidable in disordered systems. Naively, one would expect such regions to be {\it exponentially} rare: the probability to observe a thermal inclusion of size $l_T$ should correspond to $k={\cal O}(l_T)$ independent microscopic rare events, each of which occurs with probability $p$. This line of reasoning leads to a probability distribution for the length of the thermal inclusions given by 
\begin{equation}\label{Eq:rare}
p_T(l_T) \sim p^{k} \sim e^{- {c} l_T }.
\end{equation} 
Similarly, one expects that the thermal phase contains rare MBL inclusions that are exponentially distributed. These inclusions act as bottlenecks~\cite{Demler14,VHA,PVP,Reichman15,luitz2016,Znidaric16}, leading to subdiffusive transport and subballistic spreading of entanglement. However, recent RG studies suggest that thermal regions in the MBL phase are naturally fractal~\cite{ZhangHuse,Agarwal2017rareregion,GVS}: the number of independent, microscopic (UV) rare events needed to obtain a thermal region of size $l_T$ could scale only  as $k={\cal O}(l_T^{d_F})$, with $0<d_F \leq 1$. This can occur if rare {microscopic} thermal regions form a ``Cantor set''-like structure that can thermalize a whole region of size $l_T$~\cite{ZhangHuse}. The fractal structure of rare regions is physically motivated only in the MBL phase, and reflects the asymmetry of the two phases in the MBL transition: a set of measure zero of localized inclusions embedded in an otherwise thermal system cannot localize it, while fractal thermal inclusions in an otherwise localized system can be enough to thermalize it. 

The fractal structure of the thermal inclusions leads to a stretched exponential scaling of the thermal distribution. A natural ansatz for this distribution at the transition and in the MBL phase is therefore $p_T(l_T) \sim  {\rm e}^{- c \l_T^{d_F}} / l_T^{\alpha}$, where $\alpha$ and $d_F$ vary continuously in the MBL phase, and $c$ is a constant that is assumed to be regular near the transition.

At the transition point, it is necessary to have $d_F \to 0$ to recover a power-law distribution of thermal inclusions at criticality predicted by phenomenological RGs~\cite{VHA, DVP, TMdR_numerics}. In the KT scaling picture, the transition point is itself localized. This immediately constrains $\alpha_c \geq 2$, since otherwise the average size of the thermal Griffiths regions would diverge, inconsistent with localization.
Beyond this, it is natural in the KT picture that $\alpha_c = 2$, which saturates the stability bound. To see this, we note that Eq.~\eqref{eq:avalancheg} in the MBL phase implies that the effective coupling $g$ decays rapidly for $n\gtrsim n_\zeta \sim {1}/{(\zeta^{-1}  -\zeta^{-1}_c)}$. Therefore, we can identify the size of a thermal block scale as $\sim n_\zeta$.  The distribution  $p_T(l_T) \sim  {\rm e}^{- c \l_T^{d_F}} / l_T^{\alpha}$ with $d_F=0$ at criticality yields an average block size $\langle l_T\rangle \sim 1/({\alpha-2})$; linking these, we find that the transition occurs at $\alpha_c =2$. Additionally,  the simple solvable strong-disorder RG scheme for the MBL transition\cite{GVS} also predicts $\alpha_c =2$. This critical exponent implies that the average length of thermal clusters diverges logarithmically, $l^{\rm av}_T \sim \ln L$ at the transition, whereas the length of typical thermal clusters remains finite~\cite{GVS}. 

We expect $d_F$ to evolve continuously in the MBL phase, interpolating between $d_F = 0$ at the transition and $d_F =1$  deep in the MBL regime, where we recover exponentially rare thermal regions as in Eq.~(\ref{Eq:rare})~\cite{Imbrie16,ImbriePRL}. Nonetheless, there are two conceptually distinct possibilities on exactly how $d_F$ evolves in the MBL phase, which we now discuss in more detail.\footnote{A third potential scenario would involve non-fractal thermal inclusions with $d_F=1$ throughout the MBL phase, and would be hard to reconcile with a KT picture. We believe this scenario is unlikely, and we do not discuss it further in what follows.}

\subsection{Scenario (i): Fractal thermal inclusions}

Figure~\ref{fig:twotrans}(i) illustrates the first  scenario where the thermal Griffiths inclusions have continuously evolving fractal dimension. It is natural to link the evolution of $d_F$ with the scaling variable $\zeta^{-1} - \zeta^{-1}_c$, the quantity that parametrizes the position along the fixed line of Fig.~\ref{fig:KTflows}.  Given the constraints on the limiting behavior of $d_F$ discussed above, we posit that $d_F = f(\zeta^{-1} - \zeta^{-1}_c)$, where $f$ is some smooth function with $f(0) = 0$ and $\lim_{x\rightarrow\infty} f(x) = 1$, as sketched in Fig.~\ref{fig:twotrans}(i). In particular, $d_F \propto \zeta^{-1} - \zeta^{-1}_c$ near criticality.  This scenario has the advantage that it directly reconciles the line of fixed points with the exponential behavior of thermal blocks at strong disorder without any additional ingredients such as the second phase transition that is necessary in scenario (ii). However, up to this point, no numerical simulations (see Sections~\ref{sec:numerics} \& \ref{sec:numerics2}), have shown unambiguous signals of finite fractal dimensions $d_F$.

\subsection{Scenario (ii): Intermediate critical MBL phase }

\label{SecScenario2}

In the second scenario, shown in Figure~\ref{fig:twotrans}(ii), the length of the thermal blocks in the near-critical MBL phase remains power-law distributed, $p_T(l_T) \sim {l_T^{-\alpha}}$, even for $W>W_c$ --- corresponding to rare regions with $d_F=0$. The power-law scaling implies that, although the average thermal block size remains bounded in the MBL phase, there is nevertheless substantial multifractal character~\footnote{We note parenthetically that  previous studies {of microscopic models} have noted multifractal or critical behavior persisting deep into the localized phase~\cite{Rahul16,Serbyn17,2018arXiv181210283M}; it is at present unclear if these are related to our discussion here.}, indicated by the divergence of sufficiently high moments of the thermal block distribution, $\langle l_T^n \rangle_{P_T} \rightarrow \infty$ for $n>\alpha-1$~\cite{GVS}.

Given that the power-law distribution of Griffiths regions in the near-critical regime is incompatible with the (stretched) exponential behavior expected at strong disorder, one needs to reconcile the two predictions. One possible mechanism that can alter the behavior of the thermal Griffiths regions is  if some UV-scale physics that is irrelevant at the MBL-ETH critical point becomes relevant as we move along the fixed line into the MBL phase, eventually destabilizing the line of fixed points. This would then lead to a second transition deep in the MBL phase (at $W= W_c^*$); at this point, the fractal dimension $d_F$ becomes non-zero, and evolves smoothly to its limiting value of $1$ for strong disorder.  This picture, with its striking predictions of (a) an intermediate `critical MBL phase'  (that we dub, unimaginatively, `MBL$^*$') for $W_c < W<W_c^*$ and (b) a second  transition in the behavior of the thermal Griffiths regions at $W=W_c^*$, is sketched in Fig.~\ref{fig:twotrans}(ii). Possible candidates for such UV-scale physics include, for example, ``Hartree shifts'' that capture the many-body physics of l-bits and distinguish the MBL phase from the non-interacting Anderson insulator~\cite{Gopa18}; or corrections to the fully thermal behavior of thermal inclusions assumed in the avalanche picture~\cite{potirniche2018stability}. We leave a detailed analysis of this tentative second transition for future work, but we note that it could possibly be related to the breakdown of the locator expansion.
One may note the formal similarity of this scenario to the role of ${\mathbb Z}_n$ anisotropy (with $n\geq 5$) in the 2D XY model: the anisotropy is irrelevant at criticality but becomes relevant as we move along the fixed line corresponding to the power-law correlated KT phase~\cite{PhysRevB.16.1217}. 

We also note that this intermediate critical MBL* phase is reminiscent of several proposals of an intermediate phase in the literature~\cite{PhysRevLett.113.046806, PhysRevLett.115.186601, PhysRevB.94.045111,DevakulSingh}. Our work provides a clear characterization in terms of rare regions of such an intermediate phase, although our scenario is conceptually distinct  from proposals of a ``delocalized but non-ergodic'' phase, since here, the MBL* phase is fully localized. Moreover, we caution that to a large extent, these proposals are mostly based on apparent slow dynamics from small scale numerics on the \emph{thermal} side of the transition, that could also be interpreted as a finite-size critical fan.~\cite{Serbyn17}

\section{Numerical Simulations of cluster RG scheme (DVP)~\cite{DVP}\label{sec:numerics} }

In this section, we revisit the interpretation of the numerical RG of Ref.~\onlinecite{DVP} by Dumitrescu {\it et al} (DVP) in view of the proposed KT picture of the transition. This RG scheme is built on the earlier work of Ref.~\onlinecite{PVP}, but had essential changes to the treatment of interactions between resonances. In Section~\ref{sec:numericsA}, we summarize the conceptual basis of the RG and review how the avalanche phenomenon was captured. Given our previous discussion, the avalanche physics suggests that this RG would display KT scaling. However, Ref.~\onlinecite{DVP} did not consider this scenario and instead used a scaling \textit{ansatz} with a single relevant parameter. KT transitions are notoriously challenging to identify even in large-scale simulations of well-understood classical systems --- leading irrelevant corrections to scaling are suppressed only logarithmically and dominate numerical data~\cite{Weber1988MCKTlog,Kenna1997ktlog}. Therefore, in Section~\ref{sec:numericsB}, we revisit the numerical simulations with substantially better statistics. We find a power-law distribution of thermal clusters that persists deep into the MBL phase and provides evidence for KT scaling and for an intermediate critical MBL* phase [scenario (ii)].

\subsection{RG procedure}\label{sec:numericsA}

Here we highlight the conceptual construction of the numerical RG procedure. This was discussed in greater detail, including step-by-step rules, in Ref.~\onlinecite{DVP} (see also Appendix~\ref{sec:app:rulesDVP}). The RG is designed to capture the large-scale resonances between many local degrees of freedom --- called clusters --- that are characterized by only a few coarse-grained variables and can grow by interacting with other clusters. The essential physics is the competition between the typical matrix elements $\Gamma_{ij}$ that change the state of the clusters $i$ and $j$ and the typical energy mismatch $\Delta E_{ij}$ between those states.  Like other RGs for the MBL transition, it proceeds using a strong-disorder approach \cite{FisherRSRG2,FisherRSRG1, PekkerRSRGX, VHA, PVP}. Here, the strong-disorder assumption is that we may sharply distinguish interactions with $\Gamma_{ij} > \Delta E_{ij}$, that will resonantly admix clusters $i$ and $j$, from those with $\Gamma_{ij} < \Delta E_{ij}$, that only slightly dress the clusters while leaving them decoupled with respect to each other. Close to the critical point, this sharp separation can be self-consistently justified since the distribution of $g_{ij} = \Gamma_{ij}/ \Delta E_{ij}$ becomes broad. The case $g \sim 1$ will be encountered asymptotically rarely, since $g$ depends exponentially on fluctuating extensive quantities.

While the RG is a coarse-grained description, it maintains a close connection to the microscopic problem. First, the clusters are seeded at the lattice scale for a certain number of lattice sites $L$, using the uniform disorder distribution $[0,W]$ as well as the localization length $\zeta_0 \sim1/\ln(1+W^2)$ expected from the initially localized noninteracting degrees of freedom.  Second, the matrix elements are initially allowed to couple the clusters in an all-to-all manner. Although the interactions are still exponentially short-ranged in the localized basis, this correctly reflects the fact that resonances can occur between distant sites, if their energy mismatch is sufficiently small. Third, in contrast to RGs of Refs.~\onlinecite{VHA,ZhangHuse,GVS}, regions of the systems are not \textit{a priori} identifiable as thermal or insulating; instead, this is an emergent property of the distribution of clusters, depending on if they grow or not.

These RG rules are based simply on the strong disorder assumption and asymptotic form of matrix elements in the two phases. For $g < 1$, non-resonant, non-merging clusters have their direct coupling turned off $\Gamma_{ij} = 0$, capturing emerging l-bits in a locally MBL region. For $g > 1$, resonant clusters merge and are assumed to be locally fully thermal; the coupling to other clusters is given by the ETH~\cite{PhysRevE.50.888}. This procedure naturally encompasses the avalanche mechanism as one of the possible scenarios of cluster evolution~\cite{DVP}. In the RG, one may loosely think of $\Gamma_{ij}$ as setting an inverse time-scale to resonate. If $\Gamma_{ij} = 0$, clusters $i,j$ were not able to resonate on the time-scale set by their direct coupling. However, the same clusters might still resonate later, if mediated by coupling to other clusters. A sufficiently large and strongly coupled cluster can sequentially resonate with clusters which separately had no direct coupling between each other, which is the part of the avalanche process that eventually may lead to delocalization.  As such a large cluster grows, the matrix element evolution encodes an effective length-scale that plays an analogous role to $\zeta$. 

The RG terminates if the entire system thermalizes, or if no resonant bonds remain. The final state of the system is characterized by a distribution of thermal clusters of different sizes.

\begin{figure}
\includegraphics[width=\columnwidth]{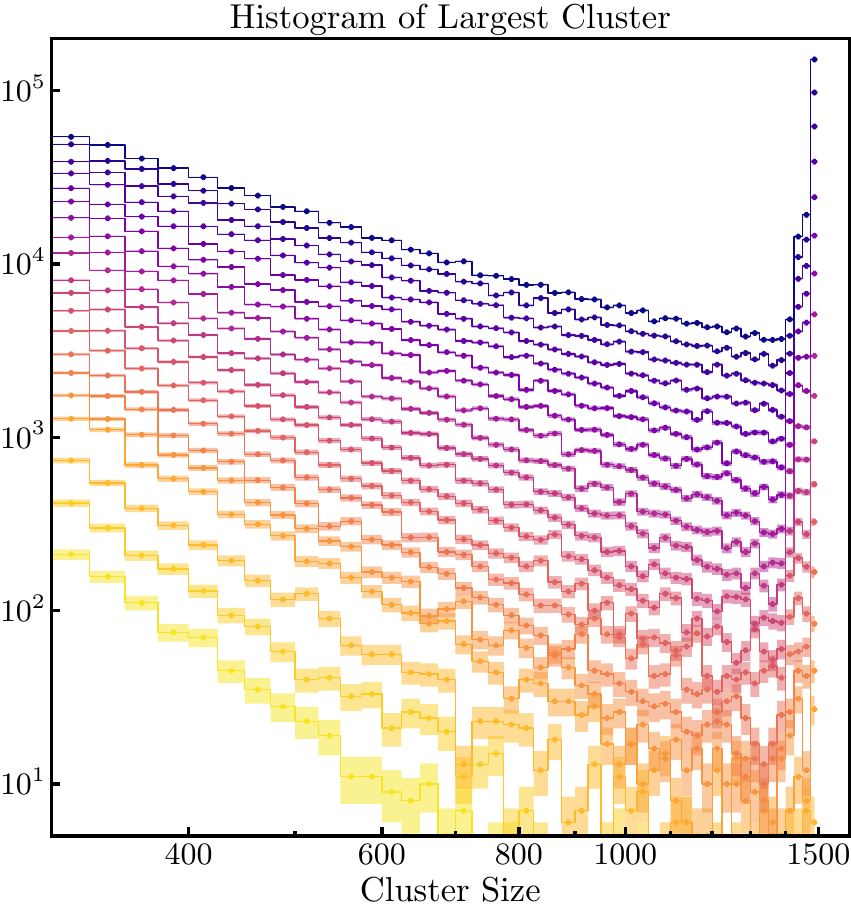}
\caption{\label{fig:histogram} DVP RG: Histograms of the number of sites (size) contained in  largest cluster for each disorder realization. The system size is $L=1500$. Different lines correspond to varying disorder bandwidths $W$: $W$ is regularly spaced from $2.04$ to $2.20$ in units of $0.01$, with three additional values $W=2.22, 2.24, 2.26$. The data fall in strict order from top to bottom (blue to gold) according to increasing $W$. Standard errors are estimated from the binomial distribution and indicated by the colored intervals around the histogram points; the bin size is $\delta x= 25$. At larger cluster sizes, the histograms show power-law behavior with a continuously varying power as well as a thermal peak close to system size $L$. The histograms are unnormalized; at small numbers of counts, noise becomes dominant.}
\end{figure}

\subsection{Numerical results}\label{sec:numericsB}

We use this RG procedure in simulations on a range of system sizes between $L=600$ and $L=4000$ initial sites with periodic boundary conditions as was done in Ref.~\onlinecite{DVP}, but with larger numbers of disorder realizations: $10^7$ (for $L \leq 1000$), $5 \cdot 10^6$ (for $L = 1500$) and $10^6$ (for $L \geq 2000$).  We tune the system across the MBL transition using the disorder strength $W$.  We analyze the resulting cluster distributions and compare them with the expectations of Sections~\ref{sec:avalancheKT} and~\ref{sec:transitions}. In particular, we will analyze both the form of the distribution at intermediate sizes as well as the fully thermalized configurations.

Throughout this section, we will consider histograms for the size of the largest cluster per disorder realization. Since we are simulating a finite size $L$, there can at most be a single cluster per disorder realization with $x > L /2$.  In fact, for sizes $x > L /2$, this distribution is identical to the distribution of all clusters.~\footnote{A cluster of size larger than $L/2$ is necessarily the largest in the sample.} Note that this distribution in a finite system is different from extreme value distributions that emerge when sampling from an infinite system. The advantage of this analysis is that each disorder realization contributes equally to the average, and therefore normalization of histograms with different $L$ can be meaningfully compared.~\footnote{For different $L$ and $W$, the number of clusters per disorder realization varies, changing the relative normalization.}

Figure~\ref{fig:histogram} shows histograms of cluster size of systems with $L=1500$. This covers a range of parameters in $W$ starting from just above the MBL transition to deep in the MBL phase.  The distribution of cluster sizes decays as a power-law and shows a finite-size `thermal peak'  of clusters of size $L$, even in the MBL phase, corresponding to a finite probability of fully thermalizing the {finite} system. 

\begin{figure*}[t]
\includegraphics[width=\textwidth]{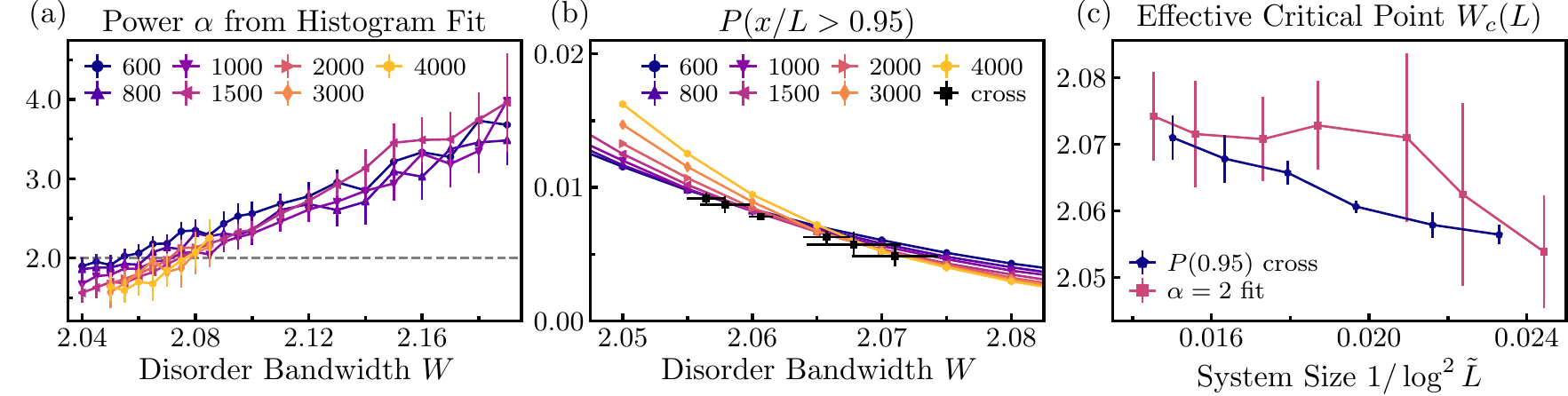}
\caption{\label{fig:powerCrossing} DVP RG: (a) Exponent $\alpha$ of a power-law fit to the distribution of the histograms of largest cluster size. The histograms are fit on the interval $[0.5L, 0.85L]$ with a bin-size of $\delta x =10$ (for $L < 1500$) or $\delta x = 50$ (otherwise). Mean values of $\alpha$ are obtained by resampling. The errors account for both the statistical error from resampling as well as the larger errors arising from varying the binning and fitting intervals. The value $\alpha_c = 2$ predicted by KT theory is shown as the grey dashed line.
(b) The fraction of disorder samples where the largest cluster size  $x \geq 0.95 L$ for various $L$.  Errors are calculated from the binomial distribution, but are smaller than the marker size. Crossing points between curves of neighboring $L$ are indicated as black crosses and serve as finite-size estimates of $W_c$. Standard errors of crossings are estimated from resampling the data.
(c) The value of effective critical disorder $W_c(L)$ slowly increases with the system size, illustrating the overestimation of the MBL phase on finite system sizes. The data is obtained from crossing points in panel (b) [$\tilde{L} = (L_1 + L_2)/ 2$ is the average length of neighboring curves] and from values of $W$ when $\alpha=2$ in panel (a) [$\tilde{L} = L$]. The finite size corrections are suppressed by a logarithm of system size, consistent with KT scaling. Errors are estimates from resampling.}
\end{figure*}

In the power-law regime, the histograms are consistent with
\begin{equation}\label{eq:powerHist}
p_\alpha(x) \propto \frac{L}{x^\alpha}.
\end{equation}
The factor $L$ takes into account the overall density {of clusters of a given size in a finite system of size $L$}; the collapse for different $L$ is shown in Fig.~\ref{fig:histall} in  Appendix~\ref{sec:app:histall}. The power $\alpha$ varies smoothly with $W$ and we show the value obtained from a least-square fit for all $L$ in Fig.~\ref{fig:powerCrossing}(a). 

The fact that the histogram remains power-law up to large values of $W$ even relatively deep in the MBL phase provides evidence for the KT {scaling picture} with scenario (ii) discussed in Section~\ref{sec:transitions}. In contrast, for a single relevant parameter, we would expect a crossover from power-law \eqref{eq:powerHist} to an exponential behavior on a scale set by the correlation length $\xi \sim (W-W_c)^{-\nu}$. Reference~\onlinecite{DVP}  analyzed the numerical data under this assumption and had difficulties extracting $\xi$ from the distribution of the cluster sizes. Reconsidering this approach, we find that the finite lower bound on $\xi$ estimated in Ref.~\onlinecite{DVP} arose from lack of statistics. In the present simulations we increased statistics by two orders of magnitude, yet we find that the power-law behavior continues down to the new noise floor (Fig.~\ref{fig:histall}).

Apart from the power-law shape of histograms, KT scaling with scenario (ii) predicts a precise form of the exponent. In particular, the analytically solvable simplified RG of Ref.~\onlinecite{GVS} predicts  $\alpha = 2 + c' \sqrt{W-W_c} + \dots$ near criticality for $W \geq W_c$. To first approximation, Fig.~\ref{fig:powerCrossing}(a) indeed shows $\alpha_c=2$ at the critical point. Below we will provide a more careful finite size scaling analysis of the numerical data, that gives further support to the critical value of $\alpha$ and KT scaling. 

The thermal peak $P_L(W)$ of the histograms itself shows scaling behavior and its finite size crossings identify the critical point~\cite{DVP}. In a simplified picture that ignores the effects of finite system size on the evolution of the RG itself, this thermal peak can be viewed as a natural consequence of binning data on a finite interval: the last bin contains the entire tail of the distribution. We therefore decompose $P_L(W) = P_\textrm{th} (W)+  \Delta P_L(W)$, 
where $ P_\textrm{th} (W) $ is the thermal fraction in the infinite-size limit and $\Delta P_L(W)$ is the contribution of all finite clusters larger than $L$. Based on consistency with one-parameter scaling, a finite value of $P_\textrm{th}({W_c})$ at  criticality was advocated in Ref.~\onlinecite{DVP}. The KT picture predicts this infinite size contribution $P_\textrm{th}({W_c}) = 0$. Nonetheless, it is consistent with a finite crossing of $P_L({W_c})$ because of the special form of $\Delta P_L({W_c})$ from Eq.~\eqref{eq:powerHist} with $\alpha_c = 2$. In particular
\begin{equation}\label{eq:KTthPeak}
P_{L}(W_c) =  \int_{L}^{\infty}p_2(x) dx \sim L \int_{ L}^{\infty} \frac{dx}{x^2} \sim {\cal O} (1),
\end{equation}
is a constant of order one and independent of $L$ up to logarithmic corrections expected from KT scaling.

Figure~\ref{fig:powerCrossing}(b) shows the fraction of samples where the largest cluster has size $x > 0.95L$.  This constitutes a measure of $P_L(W)$ but accounts for broadening due to finite size effects beyond binning. We can define an effective critical point $W_c(L)$, which takes into account finite-size drift, by considering the crossing between curves of neighboring $L$ in (b). Figure~\ref{fig:powerCrossing}(c) shows the values of $W_c(L)$ extracted in this way as well as a $W_c(L)$ from where the power-law fits of (a) have $\alpha = 2$. Within the errors set by the fitting procedures, these curves agree with one another. This agreement is entirely surprising: the thermal fraction crossing is controlled by realizations with clusters of size $L$, while the power-law fit is a property of the realizations where the largest cluster has intermediate lengths. It is, however, natural \emph{if} we assume a single power-law form of cluster distribution \eqref{eq:powerHist} and the special value of the exponent $\alpha_c = 2$ at criticality, both provided by the KT theory.  We emphasize that a simple power law with any $\alpha_c > 2$ is inconsistent with the observed crossing of $P_L(W_c)$.
 Finally, the size-dependent corrections to $W_c(L)$ are consistent with the form $\sim 1 / \ln^2 L$, expected for finite-size corrections at a conventional KT transition~\cite{Bramwell1994MagnetizationKT}.

\subsection{ Finite-size corrections and MBL* phase}

The power-law histograms of cluster size, the critical value of the exponent $\alpha_c = 2$, the consistency with the thermal fraction crossing, and the logarithmic finite-size drift discussed above provide strong support for the KT picture. Indeed, it is difficult to produce a simple alternative explanation that satisfactorily reconciles all these points. 
Nonetheless, as mentioned before, KT transitions are challenging to study in finite-size simulations, and we therefore caution that one must remain aware of the limits of interpreting the data even on systems as large as those studied here. We now discuss some of the issues related to finite-size effects.

First, the KT picture with power-law distribution \eqref{eq:powerHist} predicts the lack of a finite thermal fraction  $P_\text{th}(W_c)$. Despite Eq.~\eqref{eq:KTthPeak} giving a consistent interpretation of the data of  Fig.~\ref{fig:powerCrossing}(b), the presence of a small $P_\text{th}(W)$, which depends weakly on $W$, cannot be ruled out. Determining $P_\text{th}(W_c)$ requires a careful extrapolation of the values of $P_{L}(W_c) $ to $L\rightarrow\infty$. Regrettably,  the limited system sizes make this extrapolation poor; numerically, $P_\text{th}(W_c)$ is consistent with a large range of values, including zero and unphysical negative values (data not shown). 

Second, just as the effective critical point $W_c(L)$ has sub-leading corrections, we also expect corrections in $W$ and $L$ to the effective power $\alpha$ in Eq.~\eqref{eq:powerHist}. For example, it is not possible to meaningfully perform a scaling analysis of the exponent $\alpha(W)$ in Fig.~\ref{fig:powerCrossing}(a), to test the KT form away from the critical point. Furthermore, the finite-size corrections of $W_c(L)$ of Fig.~\ref{fig:powerCrossing}(c) are consistent with the form $\sim 1/\ln^n L$, however the available data range does not suffice to uniquely fix the power to $n=2$. Finally, if we consider integrating the cluster size distribution from various interval sizes upwards analogous to eq.~\eqref{eq:KTthPeak}, we find deviations from a constant. While expected for KT, such effects are challenging to properly interpret without analytic results for the form of leading irrelevant contributions.

The limitations imposed by the large finite size corrections are especially important with respect to the nature of the global phase diagram discussed in Section~\ref{sec:transitions}. Our numerical data are compatible with scenario (ii) of a critical MBL* phase. In order to be consistent with the finite fractal dimension of clusters deep in the MBL phase~\cite{Imbrie16}, this would indicate a second phase transition from MBL* to MBL that we did not observe in our numerics.
 Indeed, tracking the evolution of the thermal block distributions into the localized phase (Fig.~\ref{fig:histall}), we find no evidence of rounding-off of the power-law expected from an appreciably large non-zero fractal dimension of the thermal inclusions.~\footnote{{Despite the absence of curvature on a log-log scale, we note that at strong disorder, we were also able to fit the distribution of thermal clusters using a stretched exponential form. However, we caution that fitting a stretched exponential requires working with a double logarithmic scale, which is questionable even for the large systems that we can access using this numerical RG}} This remains true even for the largest disorder strength $W$ considered, where $\alpha$ is large and there is area-law entanglement scaling (cf. Ref.~\onlinecite{DVP}). As mentioned, deviations from a simple power-law form are  also expected at finite sizes due to higher-order corrections. {We caution that} if the fractal dimensions in the observed disorder window were sufficiently small,  pure power-law and stretched-exponential behaviors could not be distinguished on the accessible system sizes. 

It is entirely possible that a second transition could occur at stronger disorder within the current RG, and would just require additional orders of magnitude of statistics and system sizes to be observed. Another option is that the RG rules of Ref.~\onlinecite{DVP} are missing a necessary microscopic ingredient (see Section~\ref{SecScenario2}) to produce stretched exponential tails at strong disorder. Including such missing physics could lead to results compatible with scenario (i) of Section~\ref{sec:transitions}, where the phase is changed but the actual critical point is not, or to even more dramatic changes. We leave the resolution of this puzzle for future work, but note that at present, our numerical results favor scenario (ii) with an intermediate MBL* phase over scenario (i).

\section{Numerical Simulations of block RG scheme (VHA)~\cite{VHA} \label{sec:numerics2} }

We now consider the `block RG'  for the MBL transition originally formulated by Vosk \emph{et al.} (VHA) in Ref.~\onlinecite{VHA}. This RG was also initially interpreted assuming one-parameter scaling, with a correlation length exponent $\nu \approx 3.1$. In this section, we will present results for the near-critical localized phase on large system sizes that show consistency with KT predictions. We briefly summarize the technical steps of the RG procedure, but refer the reader to the original paper for a justification of this method.\cite{VHA} 

\subsection{RG procedure}\label{sec:numerics2A}

We begin by outlining the rules of the VHA block RG scheme (see  Appendix~\ref{sec:app:VHA} and  Ref.~\onlinecite{VHA} for details). This approach treats the system as a chain of blocks, each characterized by two parameters -- an internal rate of relaxation across  the block $\Gamma_i$,  and a typical level spacing $\Delta_i$. The dimensionless ratio $g_i  = \Gamma_i/\Delta_i$ captures the delocalized ($g_i\gg1$) or localized ($g_i\ll 1$) nature of the blocks. Adjacent blocks are connected by links, which are characterized by inter-block tunneling rates $\Gamma_{i,i+1}$.   In addition, an inter-block $g$-parameter is defined as $g_{i,i+1} = \Gamma_{i,i+1}/\Delta_{i,i+1}$, where the two-block level spacings $\Delta_{i,i+1}$ is proportional to the product of level spacings in the respective blocks.  

Initially, all the blocks have  $g_i = 1$, while the disorder  is introduced in the inter-block tunneling rates $\Gamma_{i,i+1}$. The disorder strength is parametrized by a value denoted $\ln g_0$. At any given step of RG, the link with the largest rate $\Gamma_{i,i+1}$ is decimated by merging blocks $i$ and $i+1$ into a single block. The internal relaxation rate of the new block is $\Gamma^\text{new}_i = \Gamma_{i, i+1}$. The rules for renormalizing the couplings for the new cluster to the adjacent blocks, are found by examining various limiting cases (see Appendix~\ref{sec:app:VHA}).

\subsection{Distribution of thermal regions}\label{sec:numerics2B}

\begin{figure}[t]
\centering
\includegraphics[width = \columnwidth]{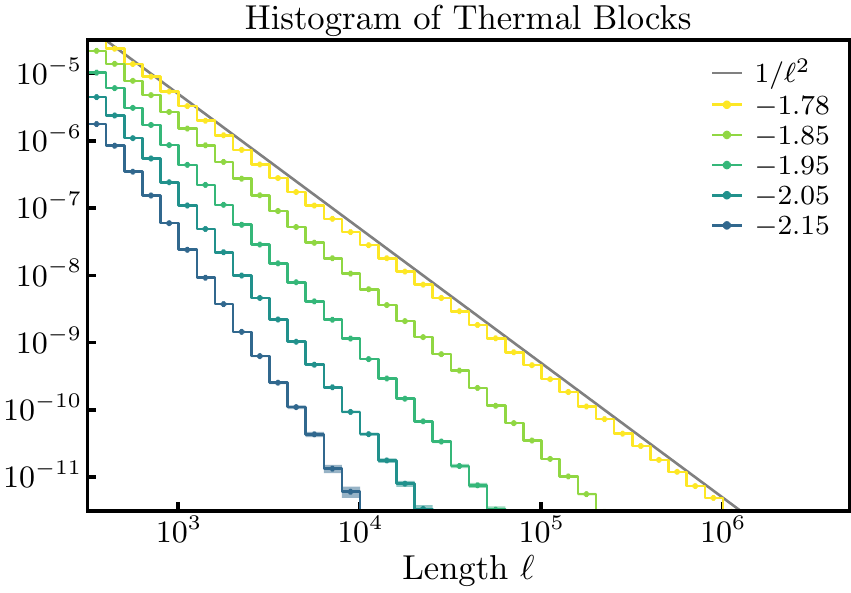}
\caption{VHA RG: Distribution of the length of thermal inclusions at criticality and into the MBL phase. The disorder strength is increased from the critical value $\ln g_0 = -1.78$ (topmost curve) up to $\ln g_0 = -2.15$ corresponding to a point deep in the MBL phase. 
At criticality the tail of the distribution  is consistent with $1/\ell^2$ behavior, and the exponent increases into the MBL phase. Error bars of the histogram were calculated based on the assumption of Poisson-distributed events in each bin of the histogram.  At criticality, we simulate systems with $L=3\cdot 10^6$ initial blocks and $2\cdot 10^3$ disorder realizations. In the MBL phase, we use $L = 5\cdot 10^6$ initial blocks and average over $4\cdot 10^4$ disorder realizations. 
\label{fig:VHA}}
\end{figure}

The VHA RG naturally delimits blocks by their $g$-value as thermal ($g_i>1$) or localized ($g_i \leq 1$). To track the distribution of thermal regions, we record the length of each newly created block $i$ whenever it crosses the threshold for thermal behavior $g_i = \Gamma_i/\Delta_i> 1$. The statistics of such lengths is accumulated over many different realizations of disorder. 

Figure~\ref{fig:VHA} shows the numerical distributions of thermal block sizes at the critical point and in the MBL regime.  Here, the critical value of the disorder strength is found to be $\ln g_0 = -1.780 \pm 0.005$ from finite-size scaling crossings in the effective eigenstate entropies~\cite{VHA} (see Appendix~\ref{sec:app:VHA}).  At criticality, we find a power-law distribution $1/\ell^\alpha$, with a power consistent with $\alpha_c = 2$. In the MBL phase near the transition, the thermal length distributions still show a power-law behavior extending over several decades, with $\alpha$ increasing for larger values of disorder strength.
Over the length-scales simulated, the distributions show slight curvature for large values of disorder. However, this curvature does not increase deeper into the MBL phase.

\begin{figure}[t]
		\centering
	    {\includegraphics[width = \columnwidth]{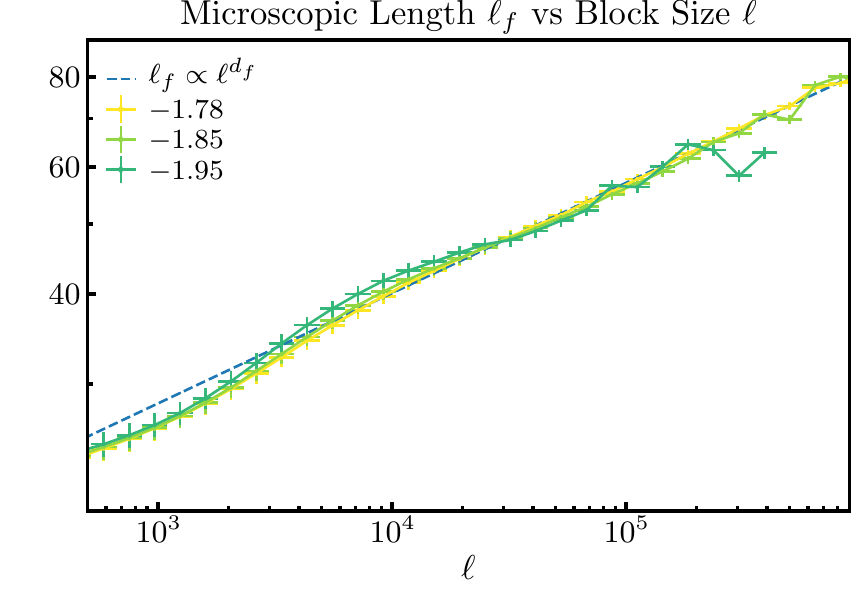}}
		\caption{ 
VHA RG: Power-law scaling between $\ell_f$ and $\ell$ allows us to extract the fractal dimension $d_F =0.153 \pm 0.005$. The fit uses the region of $\ell > 10^4$. We note that at larger value of disorder, $\ln g_0 = -1.95$, we do not see a clear power-law scaling between $\ell$ and $\ell_f$ in the same region. We use systems with $L$ ranging from $5 \cdot 10^6$ to $7\cdot 10^6$ clusters and perform averaging over $10^4$ to $1.2\cdot 10^5$ disorder realizations. The error bars correspond to the standard deviation in each bin. Since disorder is initially introduced only in the links between the blocks, there is a region of lengths where the curves for different disorder strength coincide. \label{fig:VHAF}}
\end{figure}

The clear power-law distribution of thermal inclusions at criticality with $\alpha_c = 2$, provides direct support for the KT scaling theory developed in previous sections. However, the data in Fig.~\ref{fig:VHA} is inconclusive when distinguishing between the two scenarios discussed in Section~\ref{sec:transitions}. On the one hand, we find power-law-like distributions of thermal regions away from the critical point in the MBL phase with continuously varying exponents.  This observation supports scenario (ii) of  Section~\ref{sec:transitions}, with the existence of an intermediate MBL$^{*}$ phase. On the other hand, the presence of slight curvature in $p_T(\ell)$ could be the indication of the presence of a small fractal dimension $d_F$, supporting scenario (i). However, in this latter case we would expect the curvature to increase on moving deeper into the MBL phase, which is not an  apparent trend in our simulations. Instead, the data could be also consistent with a power-law, where the power approaches its asymptotic value. 

To further address the nature of the Griffiths regions near criticality, we attempt to directly consider the internal structure of thermal inclusions to extract the fractal dimension, $d_F$. For a given thermal block $i$ of length $\ell$, we calculate the total length $\ell_f$ of those initial microscopic thermal blocks that contributed to the formation of $i$ and also have never been part of a localized block at any previous RG steps. A non-zero fractal dimension would manifest in a power-law relation $\ell_f \propto \ell^{d_F}$, with $0<d_F \leq 1$. In contrast, a fractal dimension $d_F=0$ would lead to logarithmic scaling $\ell_f \propto \ln \ell$. 

Figure~\ref{fig:VHAF}  shows the dependence of $\ell_f$ on $\ell$ at three values of disorder. The data are consistent with a small fractal dimension $d_F\approx 0.15$ both at the critical point ($\ln g_0 = -1.78$) and in the near-critical regime ($\ln g_0 = -1.85$). Deeper in the phase ($\ln g_0 = -1.95$) we do not observe a region with a clear power-law dependence between $\ell$ and $\ell_f$ and hence cannot reliably extract a fractal dimension. Note that  $d_F$ is constant within error bars at both at the critical points and over a range of disorder in the near-critical MBL phase where the exponent $\alpha$ in the thermal block distribution changes appreciably (Fig.~\ref{fig:VHA}). On the one hand, this suggests that the internal structure of the thermal blocks might be similar at criticality and in the MBL region. On the other hand,  the power law behavior at the critical point seen in Fig.~\ref{fig:VHA} implies $d_F = 0$ which contradicts the finite value of $d_F$ extracted from fits in Fig.~\ref{fig:VHAF}. Thus, we conclude that simple log-log fitting still suffer from systematic inaccuracies despite the two decades of values of $\ell$ used in these fits.

\section{\label{sec:disc} Discussion and Outlook}

In this paper, we have developed a Kosterlitz-Thouless scaling picture of the MBL transition. Ultimately the emergence of KT scaling follows from just two assumptions: that quantum avalanche processes within the localized phase drive thermalization, and the renormalization of effective matrix elements in the MBL phase due to the presence of thermal inclusions.  Taking the typical localization length (that controls these matrix elements) and the density of thermal regions as scaling variables yields a pair of RG  flow equations of the KT form. We then used various physical arguments in concert with a recently proposed, analytically solvable (but simplified) RG scheme~\cite{GVS}, to explore the implications of this KT scenario.

Our proposed KT picture of the MBL transition leads to two sharply distinct scenarios for the behavior of the near-critical localized phase. As we increase disorder to cross the transition from the thermal side, the KT picture implies that the critical point marks the beginning of a line of fixed points that continues into the MBL phase.  It is natural to  associate this line of fixed points with the continuous evolution of parameters that characterize the rare (thermal) Griffiths regions, which are expected on general grounds in disordered MBL systems. The two scenarios correspond to two distinct ways to make such a connection.  The first views the Griffiths regions as having a finite fractal dimension, and takes the line of fixed points to interpolate between fractal dimension $d_F=0$ at criticality --- this corresponds to power-law-distributed thermal regions --- to $d_F=1$ at strong disorder, as {expected} on general grounds. The second  instead takes the near-critical Griffiths regions to have vanishing fractal dimension but a continuously evolving power-law distribution. This corresponds to a phase --- dubbed MBL$^*$ --- that is distinct from the strong-disorder MBL phase, and separated from it by a second transition at which the fractal dimension becomes non-zero. Additional ideas beyond the basic assumptions needed to build the KT picture are necessary to determine which of these is realized in microscopic models of MBL.

We also turned to numerical simulations of two distinct phenomenological `cluster'~\cite{DVP} (DVP) and `block'~\cite{VHA}  (VHA) RG schemes in order to numerically test the KT scenario.  Both yield a clear power-law distribution for thermal blocks at criticality, with an exponent $\alpha_c=2$ as predicted by the KT flow. This is persuasive evidence in favor of the KT picture, given that the two RG schemes are rather different in nature.  Both show such power-law behavior extending into the near-critical MBL phase, persisting for several decades. However, they are inconclusive on the evolution of the fractal dimension $d_F$ inside the MBL phase, despite the large system sizes simulated.

A natural question raised by the present work is whether and how the RG scheme of Refs.~\onlinecite{THMdR_meanfield,TMdR_numerics} is connected to the KT scenario. We have already noted that there is a strong resemblance between the equations describing avalanches in Refs.~\onlinecite{THMdR_meanfield,TMdR_numerics} and our KT flow equations. Separately, many KT transitions have been observed in the study of disordered ground states \cite{GiamarchiSchulz,Altman2010disorderBosonsSFIns,ZYao2016SFIns1dDisorder, Doggen2017WeakStrong1dBoseGlass}, although the physics in those cases are different. Exploring the connections between the KT scaling of the highly excited states in the MBL phase presented here and in the ground state is an open direction.~\cite{Dupont2019mblGroundState}
Another obvious direction along which our work can be developed is to build in more microscopic input. The link that we have provided between the analytically solvable RG of Ref.~\onlinecite{GVS} and the microscopically motivated models of Ref.~\onlinecite{DVP} has already allowed us to connect the KT scenario to concrete physical observables. Going beyond this to truly microscopic models requires more effort. Although traditional techniques such as finite-size scaling collapses are likely challenging in light of our KT scenario, other observables may distinguish KT from other scaling behavior. It would be interesting to understand what the analogue quantity of the universal Luttinger liquid parameter of disordered ground state KT transitions \cite{GiamarchiSchulz} is and the relationship to the universal exponent $\alpha_c = 2$ presented in this paper. We also note, that a very recent exact diagonalization results appear to be consistent with the KT prediction of algebraic behavior of thermal cluster distributions within the MBL phase, with an exponent $\alpha_c  \approx 2$ at criticality~\cite{JensPaper}.

Finally, we note that the link with the behavior of thermal Griffiths regions suggests that it {is}  possible, at least in principle, to extract this exponent experimentally. We can imagine a cold atom setup where the system is quenched into the MBL regime from an initial configuration with a local `imbalance' (period-two density modulation)~\cite{schreiber2015observation}.  As it does not correspond to a conserved quantity, the imbalance will persist to long times in the localized regions, but  decay away in the thermal blocks. It should be {feasible} to measure spatial fluctuations of the imbalance using a quantum gas microscope~\cite{Bakr:2009aa}. This yields the distribution of   thermal block sizes, allowing us to extract $\alpha$, both at criticality, as well as into the localized phase. Such a measurement, performed with enough sensitivity, may also be able to distinguish between the two scenarios proposed for the nature of the near-critical phase.

\acknowledgements
We thank S.~Gopalakrishnan, D.A.~Huse, V.~Khemani, A.~Morningstar, W.~De Roeck, and S.~Roy for discussions and in particular A.C.~Potter for previous collaborations on this topic. We would like to especially thank D.A.~Huse and A.~Morningstar for multiple discussions and suggestions related to VHA, and for sharing their unpublished results on a generalization of the RG of Ref.~\onlinecite{GVS}. The Flatiron Institute is a division of the Simons Foundation.  AG acknowledges support of the Swiss National Science Foundation. This work was supported by the US Department of Energy, Office of Science, Basic Energy Sciences, under Award No.~{DE}-{SC0019168} (RV). We are grateful to the KITP, which is supported by the National Science Foundation under Grant No. NSF PHY-1748958, and the  program ``The Dynamics of Quantum Information'', where part of this work was completed.

\begin{appendix}

\section{DVP Cluster RG procedure }\label{sec:app:rulesDVP}

As stated in the main text, the DVP RG rules used in Section~\ref{sec:numerics} are identical to those introduced in Ref.~\onlinecite{DVP}. Here we repeat the step-by-step RG procedure. 

The elementary object of the RG is a cluster $i$, characterized by its bandwidth $\Lambda_i$ and the number $n_i$ of effective `spins-1/2' contained within it. The cluster many-body level spacing  is $\delta_i = \Lambda_i / (2^{n_i} - 1)$. We introduce matrices $\Gamma_{ij}$ and $\Delta E_{ij}$ that capture, respectively, the matrix elements of transitions between different clusters and the energy mismatch between the clusters.

We initialize a single disorder realization of the RG with a chain of $L$ clusters each containing a single `spin-1/2' degree of freedom ($n_i = 1$). We draw a set of $L$ random fields $\mu_i$ from a uniform distribution on $[0, W]$; this determines the initial bandwidth of each cluster $\Lambda_i = \mu_i$ and the energy mismatch $\Delta E_{ij} = \vert \mu_i - \mu_j \vert$. The initial matrix elements are initialized as $\Gamma_{ij} = V \left( e^{-\vert i - j \vert / \zeta_0} + e^{- \left\vert L - \vert i - j \vert \right\vert  / \zeta_0} \right)$, with $\zeta_0 = {2}/{\ln(1 + W^2)}$. The two terms in $\Gamma$ reflect periodic boundary conditions; we always set $V = 0.3$. {We note that the disorder strength} $W$ enters both the local bandwidth at the UV scale and the range of effective interaction between clusters.

A single step of the RG has three elements:
\begin{enumerate}
\item \emph{Identify resonant clusters and merge:} clusters connected by a path of resonant bonds ($g_{ij} = \Gamma_{ij} / \Delta E_{ij} \geq 1$) are merged into a  new resonant cluster $\{k \} \to K$. The  number of spin-1/2 in the new cluster is $n_K = \sum_{i \in \{k\} } n_i$. The new bandwidth is
$\Lambda_K = \sqrt{\sum_{i \in \{k\} } \Lambda_i^2 + \sum_{i, j \in \{k\} } \Gamma_{ij}^2 }$,
and the level spacing is updated from its definition~$\delta_K= \Lambda_K / (2^{n_K} - 1)$.

\item \emph{Update matrix elements:} Consider two sets of resonant clusters, which are separately merging $\{i\} \to I$ and $\{j\} \to J$. The coupling between the resulting clusters $I,J$ is set by ETH to be
\begin{equation*}
\Gamma_{IJ} = \left[\max_{\substack{i' \in \{i\},\\ j'\in\{j\}}}\Gamma_{i'j'}\right] e^{-(n_{I} +n_{J} - n_{i'} -n_{j'})s_{\text{th}}/2}, 
\end{equation*}
where  $s_{\text{th}}=\ln 2$ is the thermal entropy density. This form also applies when only one cluster changes.

If two clusters were \emph{both} unchanged during a RG step, the coupling is turned off:
\begin{equation*}
\Gamma_{IJ} = 0,
\end{equation*}
reflecting the emergence of decoupled l-bits.

\item \emph{Update energy mismatch:}
\begin{equation*}
\Delta E_{IJ} = \begin{cases} 
\max(\delta_I - \Lambda_J, \delta_J), & \Lambda_I \geq \delta_I \geq \Lambda_J \geq \delta_J,  \\ 
\max(\delta_J - \Lambda_I, \delta_I), & \Lambda_J \geq \delta_J \geq \Lambda_I \geq \delta_I,  \\ 
{\delta_I \delta_J}/{\min(\Lambda_I, \Lambda_J)} & \mathrm{otherwise}. \end{cases}
\end{equation*}
The first two cases are invoked when the level spacing of one cluster exceeds the bandwidth of the other, and generally occur only in the first RG steps.
\end{enumerate}

RG steps are iterated until there are no resonant couplings remaining or the entire system is thermalized into a single cluster.

\section{VHA Block RG procedure \label{sec:app:VHA}}

For the sake of completeness, we describe the ``VHA'' RG procedure.\cite{VHA}
The system is initialized as a chain of blocks consisting of spin-$1/2$ objects. Each block is characterized by an internal tunneling rate $\Gamma_i$ and a typical level spacing $\Delta_i = W/2^L$, where $W$ is a bandwidth and $L = \ln_2 (100)$ is a length of the block. Initially all the blocks have identical value of $g_i \equiv \Gamma_i/\Delta_i = 1$. The coupling between the adjacent blocks is taken into account via inter-block tunneling rates $\Gamma_{i,i+1}$. One also defines an effective two-block level spacing as $\Delta_{i,i+1} = W/ 2^{L_i + L_{i+1}}$ and a coupling $g_{i, i+1} = \Gamma_{i,i+1}/\Delta_{i,i+1}$. 

By construction, the tunneling rate through the link is smaller than the intra-block tunneling rate. The disorder is introduced into the system via random initial values of link parameters $g_{i,i+1}$.  In order to incorporate the above constraint, one first defines the auxiliary couplings $\tilde{g}_{i,i+1}$ that are drawn from the standard log-normal distribution with the mean $\ln g_0$ and variance set to one. The parameter $\ln g_0$ is used as the tuning parameter for the MBL transition in the VHA RG. The actual couplings are defined as $g_{i,i+1} = 1/(\tau_{i,i+1}\Delta_{i,i+1})$, where $\tau_{i,i+1} = 1/(\tilde{g}_{i,i+1}\Delta_{i,i+1}) + 2/\Gamma_0$ and $\Gamma_0$ is the intra-block rate at the beginning of the procedure. 

At subsequent steps of the RG the ``fastest'' link, i.e. with the largest value of $\Gamma_{i,i+1}$ is decimated, so that the blocks it connects form a new block. This requires the renormalization of the adjacent links to other blocks. We briefly clarify and summarize the rules below, referring the reader to Ref.~\onlinecite{VHA} for their justification. Denoting the link with the largest rate, $\Gamma_{12}$, to be between blocks 1 and 2, the link to the right transforms as $\Gamma_{23}\to\Gamma_R$ (the left link is defined analogously). The value of the  $\Gamma_R$ is determined according to two possible rules,
\begin{eqnarray}\label{eq:app:VHA2}
\Gamma_R &=& \frac{\Gamma_{12} \Gamma_{23}}{\Gamma_2}, \\
\label{eq:app:VHA}
\frac{1}{\Gamma_R} &=& \frac{1}{\Gamma_{12}} + \frac{1}{\Gamma_{23}} - \frac{1}{\Gamma_2}.
\end{eqnarray}
The rule being applied depends on the value of $g$ parameters, and is summarized in Table~\ref{TabVHA}.
\begin{table}[t]
\center
\begin{tabular}{|l|c|c|}
 \hline
        ~ &$ g_{12} \ll 1$ & $ g_{12} \gg 1$ \\
       \hline
      $g_{23}\ll 1$ & \eqref{eq:app:VHA2}&\eqref{eq:app:VHA} if $g_1\ \&\ g_2 \gg 1$  \\
      \,&\,&\eqref{eq:app:VHA2} if $g_1$ or $g_2 \ll 1$\\ 
      \hline
     $g_{23}\gg 1$& \eqref{eq:app:VHA} if $g_2\ \&\ g_3 \gg 1$  & \eqref{eq:app:VHA}\\
     \,&\eqref{eq:app:VHA2} if $g_2$ or $g_3 \ll 1$&\\
      \hline
 \end{tabular}
\caption{Rules of the VHA RG.\label{TabVHA}}
\end{table}
The same procedure is also applied to the link to the left of the merged blocks (if such link exists, since we are using the open boundary conditions). The decimation procedure described above is repeated until only single block remains. 

\begin{figure}[b]
\includegraphics[width=\columnwidth]{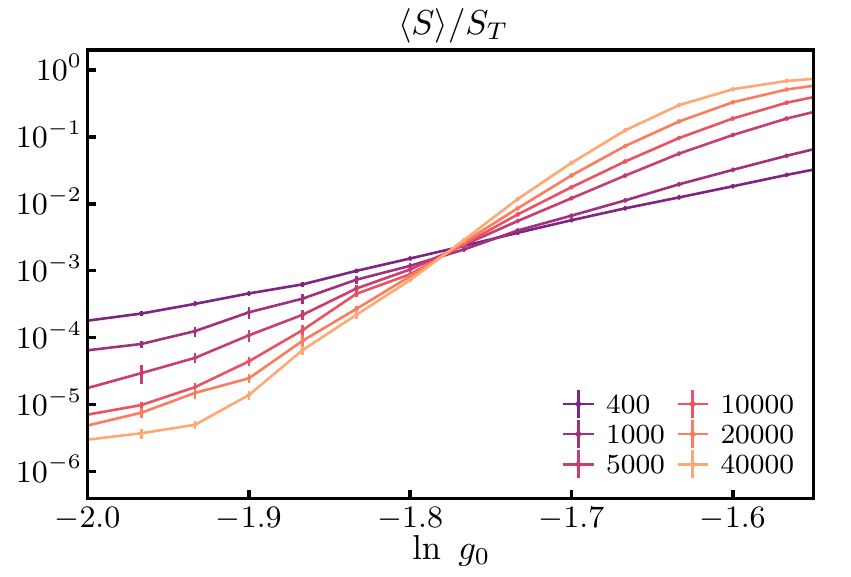}
\caption{\label{fig:app:VHA} Finite size scaling of normalized average entanglement entropy\cite{VHA} shows a crossing at a particular disorder strength, $\ln g_0 = -1.78\pm 0.005$. For stronger disorder $\ln g<\ln g_0$ entropy decreases, corresponding to the MBL phase, whereas for $\ln g>\ln g_0$ normalized entropy tends to the thermal value upon increasing system size. The number of disorder realizations used is $10^4$ for $L=400; 1000$,  $2\cdot 10^4$ for $L = 5000,10000$, and $3\cdot 10^4$ for $L=20000; 40000$. }
\end{figure}

As our goal is to study the broadly distributed thermal block distribution of the RG, it is crucial to study large system sizes averaged over many disorder realizations. This requires an efficient numerical implementation of the RG. The VHA procedure is based on the decimation of the largest coupling $\Gamma_{i,i+1}$ and renormalization of couplings to its two spatial neighbors. A na\"ive search for the block with the largest coupling in a list requires $\mathcal{O}(L)$ operations at each RG step and is prohibitively inefficient. Instead, we use a container that allows access via two distinct, \emph{sorted} labels -- the spatial position of the block and the coupling $\Gamma_{i,i+1}$ to the next block on the right. Each RG step now costs $\mathcal{O}(1)$ operations to search and $\mathcal{O}(\log L)$ to resort the energy index after merging and renormalization. We use the \texttt{boost.multiindex} container from the C++ Boost libraries \cite{boostlib:mi} for this strategy.

\begin{figure}[b]
\includegraphics[width=\columnwidth]{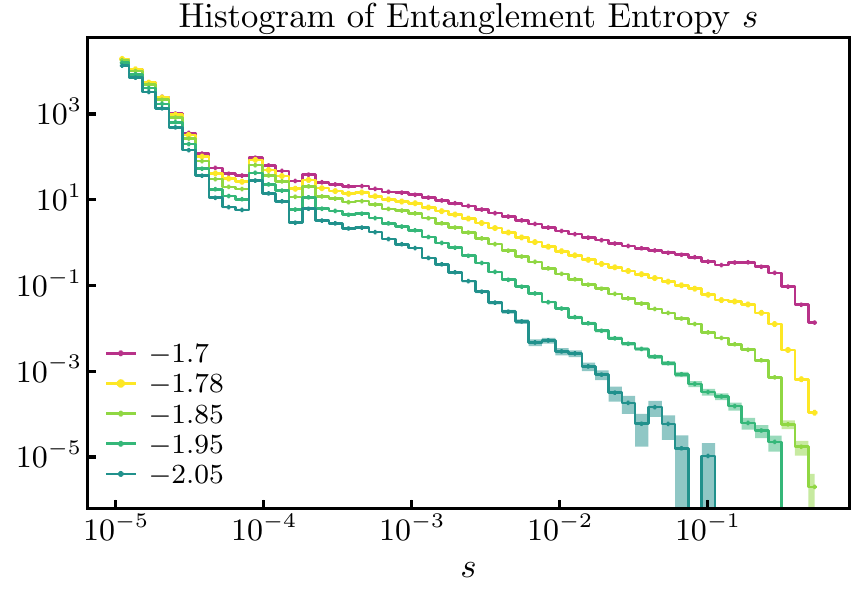}
\caption{Distribution of normalized effective entanglement entropy $s = \langle S\rangle/S_T$ has a broad tail at the MBL transition and into the localized phase. In the delocalized phase (topmost curve, $\ln g_0 = -1.7$) the distribution is non-monotonic, suggesting the formation of a maximum near thermal value of entanglement. In contrast, at the transition ($\ln g_0 = -1.78$) and in the MBL phase we observe a power-law-like tail that extends to volume-law values of $s\sim 0.2$. The data was obtained on small system sizes of $L=10^4$ blocks and $10^6$ disorder realizations.}
\label{fig:entropy_distr_VHA}
\end{figure}

Because of the large system sizes simulated, our code stores the logarithms of couplings $\Gamma$ to avoid numerical overflow. 
While Eq.~(\ref{eq:app:VHA2}) is naturally formulated via logarithms, we need to regularize  Eq.~(\ref{eq:app:VHA}) to address when one of the $\Gamma$s in the denominator becomes too small. 
We observe that the RG procedure is organized in a way preserving $\Gamma_{ij} < \Gamma_k$ for any $i,j,k$ at each step of the RG. Due to the relation $\Gamma_2 > \Gamma_{12} > \Gamma_{23}$ and a strong-randomness character of the scheme, the main contribution to $\Gamma_R$ comes from $\Gamma_{23}$. Hence, we apply the rule \eqref{eq:app:VHA} unmodified when $\ln\Gamma_{23}\geq \lambda_c = -200$. When coupling $\ln\Gamma_{23} <\lambda_c$ we assume  $\Gamma_R = \Gamma_{23}$. We check that our results do not depend on the value of the cut-off $\lambda_c$ when it is chosen to be small enough.

In order to define the critical disorder strength of the transition, we exactly follow the recipe of the original Ref.~\onlinecite{VHA} which defined a measure of the effective entanglement entropy of the eigenstates. The analogue of bipartite entanglement entropy is defined as the logarithm of the number of product states needed to represent true eigenstates of the two blocks coupled by the central link. The number of such product states is approximated by the ratio of coupling to the level spacing, $g_{i,i+1} = \Gamma_{i,i+1}/\Delta_{i,i+1}$. Using this motivation, Ref.~\onlinecite{VHA} defines the entanglement as $S = \ln (1+g_\text{max})$. Here parameter $g_\text{max}$ is the maximal value taken by $g_{i, i+1}$ for the block $i$ that spans the middle of the initial system over the entire flow of the RG.

We normalize the `entanglement' $S$ defined above by the value of a thermal entropy, $S_T = L\ell_0\ln 2$, where $L$ is the number clusters each having microscopic size $\ell_0$. The normalized entanglement, $S/S_T$ approaches one in the ergodic phase and zero in the MBL phase. We reproduce the results of Ref.~\onlinecite{VHA} in Fig.~\ref{fig:app:VHA} and find the critical disorder strength $\ln g_0 = -1.78\pm 0.005$, slightly smaller than in the original work ($\ln g_0 = -1.73$).

Finally, we examine the distribution of the normalized entanglement $s=S/S_T$ across the MBL transition in Fig.~\ref{fig:entropy_distr_VHA}.
At critical disorder we find a broad tail of $p(s)$ that has parts consistent with power-law scaling $p(s)\sim 1/s$ that slightly differs from the $p(s)\sim 1/s^{0.9}$ previously reported, presumably due to the slightly increased value of critical disorder. A more surprising finding is the persistence of a broad tail of $p(s)$ away from the critical point and extending into the MBL phase. This tail, which we are able to access because of the much larger system sizes and better statistics, emerges \emph{in addition} to the exponential decay of  $p(s)$ for small values of $s$ reported in Ref.~\onlinecite{VHA}. The existence of such broad tails could be consistent with our KT picture.

\begin{figure*}
\section{\label{sec:app:histall} Histograms for various $L$ and $W$ (DVP)} 
\begin{centering}
\includegraphics[width=0.96\textwidth]{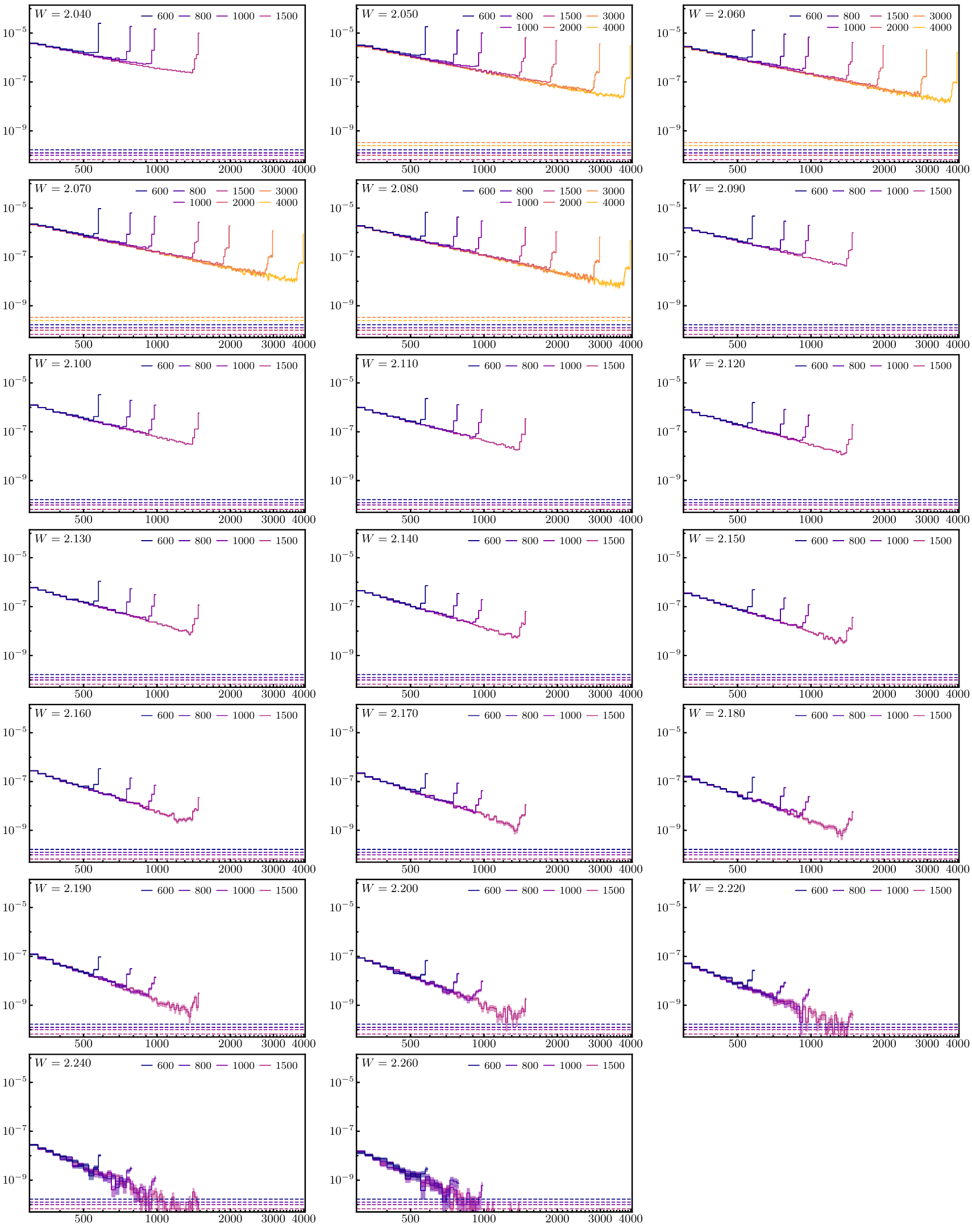}
\end{centering}
\caption{Histograms of the number of sites (size) contained in the largest cluster for each disorder distributions with different system sizes $L$ using the DVP RG.  The histograms are normalized both by the number of disorder realizations and the system size $L$. Standard errors are indicated by the colored intervals; the bin size is $\delta x= 25$. For each dataset, the limit where there is only a single count $1 /(N_{\textrm{dis} }L)$  is shown by dashed lines.}
\label{fig:histall}
\end{figure*}

\end{appendix}
\
\newpage

\bibliography{MBL}

\end{document}